\documentclass[aps,prl,twocolumn,nofootinbib,superscriptaddress,showpacs,preprintnumbers,notitlepage]{revtex4-1}
\usepackage[utf8]{inputenc}
\usepackage{amsmath,amsfonts,amssymb}
\usepackage{braket}
\usepackage{graphicx}
\usepackage{bm}
\usepackage{placeins}
\usepackage{xfrac}
\usepackage[capitalize]{cleveref}
\usepackage[mode=buildnew]{standalone}

\begin{document}
\title{Dynamical large deviations of two-dimensional kinetically constrained models using a neural-network state ansatz}
\author{Corneel Casert}
\email{corneel.casert@ugent.be}
\affiliation{Department of Physics and Astronomy, Ghent University, 9000 Ghent, Belgium}
\author{Tom Vieijra}
\affiliation{Department of Physics and Astronomy, Ghent University, 9000 Ghent, Belgium}
\author{Stephen Whitelam}
\affiliation{Molecular Foundry, Lawrence Berkeley National Laboratory, 1 Cyclotron Road, Berkeley, California 94720, USA}
\author{Isaac Tamblyn}
\email{isaac.tamblyn@nrc.ca}
\affiliation{National Research Council Canada,
   Ottawa, Ontario, Canada}
\affiliation{Vector Institute for Artificial Intelligence,
  Toronto, Ontario, Canada}

\begin{abstract}
We use a neural network ansatz originally designed for the variational optimization of quantum systems to study dynamical large deviations in classical ones. We obtain the scaled cumulant-generating function for the dynamical activity of the Fredrickson-Andersen model, a prototypical kinetically constrained model, in one and two dimensions, and present the first size-scaling analysis of the dynamical activity in two dimensions.
These results provide a new route to the study of dynamical large-deviation functions, and highlight the broad applicability of the neural-network state ansatz across domains in physics.
\end{abstract}
\maketitle

\textit{Introduction---} Dynamical systems, which include glassy~\cite{Garrahan2007DynamicalGlasses,Garrahan2009First-orderHistories, Garrahan2018AspectsDynamics}, driven~\cite{Vaikuntanathan2014DynamicNetworks, Visco2006FluctuationsGases, Bunin2012LargeBehavior, Mehl2008LargeSystems, Derrida2003ExactProcess}, and biochemical systems~\cite{Seifert2012StochasticMachines,McGrath2017BiochemicalWork}, are defined by ensembles of stochastic trajectories, much as equilibrium systems are defined by ensembles of configurations.
Trajectories can be characterized by time-extensive trajectory observables, such as  dynamical activity~\cite{Garrahan2007DynamicalGlasses, Garrahan2009First-orderHistories, Jack2014LargeTrajectories}, entropy production~\cite{Seifert2005EntropyTheorem, Maes2017FreneticProduction}, or other currents~\cite{Bodineau2007CumulantsStates, Lecomte2010CurrentEquilibrium, Gingrich2016DissipationFluctuations}.
The fluctuations of these observables are often described by large-deviation functions---the scaled cumulant-generating function (SCGF) and the rate function---which play a role analogous to thermodynamic potentials for equilibrium systems~\cite{Touchette2009TheMechanics, Touchette2011LargeSystems}. 
Calculating large-deviation functions is a challenging task, requiring the use of advanced methods based on e.g. cloning~\cite{Giardina2006DirectFunctions, Lecomte2007ATime, Nemoto2016Population-dynamicsControl}, or the use of guiding or auxiliary dynamics~\cite{Ray2018ExactDynamics, Jacobson2019DirectAnsatz, Ray2020ConstructingMinimization}. 
Recently, neural networks have been used to construct such auxiliary dynamics~\cite{Whitelam2019EvolutionaryDeviations, Oakes2020ADeviations, Rose2020ASampling}.

Here we demonstrate the ability of the neural-network state ansatz, which shall represent the long-time configurational probability distributions associated with rare trajectories, to calculate  the large-deviation functions of dynamical systems in one and two dimensions, inspired by their recent success in the variational optimization of quantum systems~\cite{Carleo2017SolvingNetworks}.
The similarities between variational energy minimization in quantum systems and finding the SCGF as the largest eigenvalue of a tilted generator have inspired variational techniques for studying large deviations in dynamical systems, in particular tensor network methods ~\cite{Gorissen2009Density-matrixTransitions, Banuls2019UsingModels, Helms2019DynamicalStates, Helms2020DynamicalNetworks}.
However, while current variational approaches to calculating large-deviation functions are usually limited to one-dimensional systems, the flexibility of the neural-network ansatz allows for straightforward generalization to higher spatial dimensions. 
We calculate the large-deviation functions for dynamical activity in a prototypical model of slow dynamics, the Fredrickson-Andersen (FA) model, in one and two dimensions, and present the first size-scaling analysis of the dynamical activity in two dimensions.
Although we shall here focus our study on the FA model, our method of obtaining large-deviation functions as described below is  very generally applicable.
The ease of extension of this approach to two dimensions opens new avenues for the efficient study of dynamical large deviations, and demonstrates the broad applicability of the neural-network state ansatz to classical dynamical problems.\\

\begin{figure*}[t]
\includegraphics[width=\textwidth]{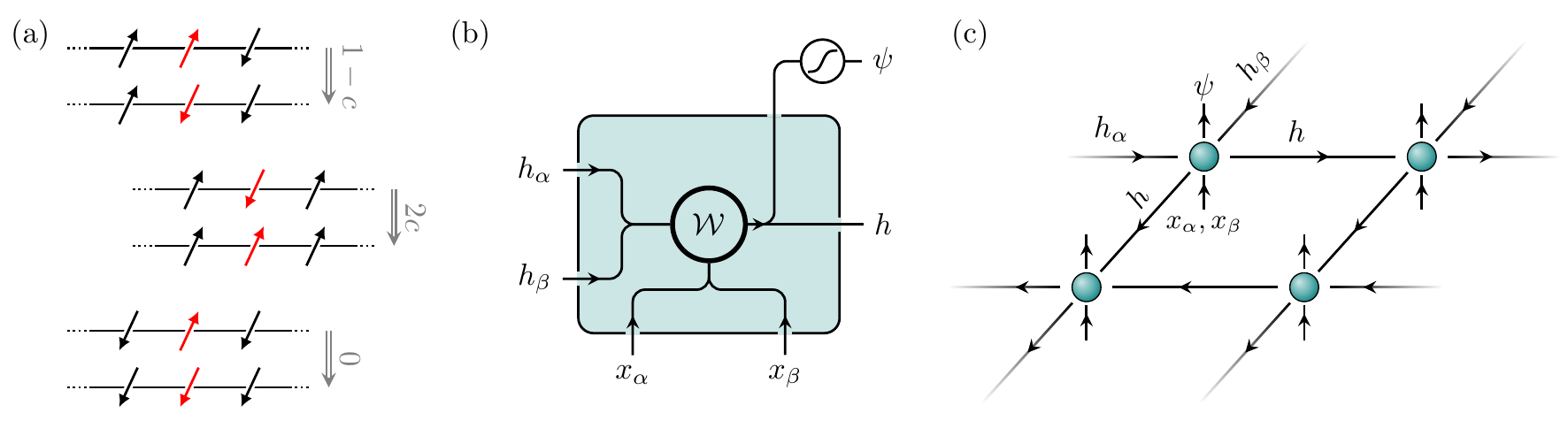} 
\caption{(a) Example transition rates of the Fredrickson-Andersen model studied in this work for the one-dimensional case. (b) For a two-dimensional system, an RNN cell with learnable parameters $\mathcal{W}$ calculates a new hidden state $h$, given the hidden and visible states of previous lattice sites $\alpha$ and $\beta$, and passes this on to other lattice sites. The hidden state is further processed to calculate the normalized conditional probability amplitude $\psi(x_i|x_{j<i})$.
(c) This RNN cell is applied to each site of a two-dimensional lattice, and the RNN traverses the lattice row by row in a zig-zag path to calculate the total probability amplitude of a configuration. Note that the hidden state is passed on in both the vertical and horizontal direction, respecting the geometry of the system under study. This probability amplitude is then used for the variational optimization of the scaled cumulant-generating function, $\theta(s)$. }
\label{fig:rnn}
\end{figure*}

\textit{Model and observables---} The FA model consists of a lattice of $N$ binary spins $i=1,\dots,N$, which take values $n_i=1$ (up) or $n_i=0$ (down).
Spin $i$ flips up (resp. down) with rate $f_i c$ (resp. $f_i(1-c)$), where $c$ is a parameter and $f_i=\sum_{j\in\text{nn}(i)}n_j$ the number of nearest-neighbor up-spins. 
The latter factor is a kinetic constraint that renders the dynamics of the model slow or glassy~\cite{Fredrickson1984KineticTransition, Butler1991TheModel,Ritort2003GlassyModels, Garrahan2010KineticallyModels}. 
This dynamics (Fig.~\ref{fig:rnn}a) is described by the generator
\begin{equation} \label{eq:W}
    W = \sum_i  f_i\, [c(\sigma_i^++n_i-1) + (1-c)(\sigma_i^- - n_i)],
\end{equation}
where $\sigma_i^\pm$ flips site $i$ up or down. 
We work with open boundary conditions by connecting each spin on the boundary of the lattice to a site in the down state. 
The configuration with all sites in the down state is disconnected from the rest of configuration space due to the kinetic constraints; we only consider dynamics without this configuration.

We will study the large deviation properties of the (intensive) dynamical activity $k=K/t$ of the FA model in one and two dimensions.
The activity of trajectory $\omega$ of length $t$ is equal to $K(\omega)$, the number of configuration changes in the trajectory. 
The probability distribution for activity in the FA model adopts for long times the large-deviation form $P(K) \approx e^{-tJ(k)}$, where the rate function $J(k)$ quantifies the likelihood of observing atypical values of activity~\cite{Touchette2009TheMechanics}.
Information equivalent to that  contained in $J(k)$ can be obtained from its Legendre transform, the scaled cumulant-generating function (SCGF), $\theta(s) =-\min_k\left(sk+J(k)\right)$.

The SCGF can be obtained as the largest eigenvalue of a modified or tilted generator, whose matrix elements connecting microstates $x$ and $y$ are
\begin{equation}\label{eq:tilted}
W^s_{xy} = W_{xy}e^{-s}(1-\delta_{xy}) - R_x\delta_{xy}.
\end{equation}
Here $W_{xy}$ are the matrix elements of the original generator, in this case Eq.~(\ref{eq:W}), and $R_x= \sum_{y\neq x}W_{xy}$~\cite{Lebowitz1999ADynamics, Touchette2009TheMechanics, Derrida2019LargeEquation}.
The SCGF can therefore be obtained by solving the eigenproblem $W^s|P^s\rangle = \theta(s) |P^s\rangle$, where the right eigenvectors $|P^s\rangle$ contain the configurational probabilities in the long-time limit for trajectories conditioned on \mbox{$\left<k\right> = \text{d}\theta(s)/\text{d}s$}.

The large-deviation properties of the one-dimensional version of the FA model are well-studied and reveal a dynamical phase transition between an active and inactive phase at a size-dependent value of $s$~\cite{Garrahan2009First-orderHistories, Banuls2019UsingModels, Nemoto2017Finite-SizeModel, Bodineau2011ActivityDynamics, Bodineau2012FiniteModel}, marked by a singularity in the SCGF. We show that a neural-network state ansatz can yield comparable results. We then use it to provide the first finite-size scaling analysis of the FA model in two dimensions.\\

\textit{Recurrent neural-network states---} Artificial neural networks (ANN) can be used as a variational ansatz by mapping configurations $\bm{x} \equiv (x_1,\ldots,x_N)$ of an $N$-site lattice system to their corresponding probability amplitude $\psi(\bm{x})$, which defines the state $|\psi\rangle = \sum_{\bm{x}} \psi(\bm{x})|\bm{x}\rangle$.
This ansatz has recently been shown to be capable of faithfully representing highly entangled quantum systems~\cite{Carleo2017SolvingNetworks, Nagy2019VariationalSystems, Hartmann2019Neural-NetworkDynamics, Vicentini2019VariationalSystems, Yoshioka2019ConstructingSystems,Choo2018SymmetriesStates,Choo2019Two-dimensionalStates, Vieijra2020RestrictedSymmetries,Melko2019RestrictedPhysics, Pilati2019Self-learningMachines, Ferrari2019NeuralFunctions, Sehayek2019LearnabilityMachines, Westerhout2020GeneralizationStates, Szabo2020NeuralProblem, Nomura2017RestrictedSystems,Deng2017MachineStates, Deng2017QuantumStates, Carleo2018ConstructingNetworks, Sharir2019DeepSystems,Hibat-Allah2020RecurrentFunctions,Roth2020IterativeNetworks}, and has found use in quantum state tomography~\cite{Torlai2018LatentOperators, Torlai2018Neural-networkTomography,Torlai2019IntegratingReconstruction, Carrasquilla2019ReconstructingModels}.
The expressivity of the ANN ansatz depends on the architecture of the neural network, and typical choices include restricted Boltzmann machines, fully-connected and convolutional neural networks, and autoregressive neural networks.
Here we use autoregressive neural networks, a popular architectural choice for complex machine learning tasks such as natural language processing, sequence generation or handwriting recognition~\cite{Chung2014EmpiricalModeling, Graves2012SupervisedLabelling, Graves2013GeneratingNetworks, Graves2013SpeechNetworks, VanDenOord2016PixelNetworks}.
A state defined by such a network can be sampled in parallel without Markov chains, which is particularly useful for physical regimes where Markov chains struggle to propose uncorrelated configurations, and allows for the efficient use of state-of-the-art computing infrastructure such as massively parallel graphical processing units.
Examples of autoregressive neural networks include PixelCNN~\cite{Sharir2019DeepSystems} and recurrent neural networks (RNN)~\cite{Hibat-Allah2020RecurrentFunctions, Roth2020IterativeNetworks}. 
We use the RNN ansatz of~\cite{Hibat-Allah2020RecurrentFunctions, Roth2020IterativeNetworks}, which was shown to be highly efficient in the optimization of quantum systems.
An RNN is defined by its elementary building block, the RNN cell, which is a parametrized non-linear function that sweeps over the lattice site by site.
For a given one-dimensional configuration $\bm{x}$, at each lattice site $i$ the RNN cell receives the ``visible'' state $x_{i-1}$ from the previous site, as well as the ``hidden'' state $h_{i-1}$, which contains information from the previously encountered degrees of freedom and serves as a form of memory.
From this, the RNN cell calculates the hidden state for the current lattice site, $h_i$,  and further processes this hidden state to obtain a conditional probability amplitude $\psi(x_i|x_{i-1},\ldots,x_1)$ for a new visible state depending entirely on $x_{j<i}$ encountered earlier on the lattice.
A new visible state is obtained by sampling $x_i$ from the distribution $P(x_i|x_{i-1},\ldots,x_1)=\left|\psi(x_i|x_{i-1},\ldots,x_1)\right|^2$ which, together with the new hidden state, can be used as input for the next site.
Starting from an initial visible and hidden state, the whole lattice can be traversed in this way, which allows for highly parallel sampling and calculation of probability amplitudes.
This procedure can be naturally extended to higher dimensions, e.g. for a two-dimensional system, we provide the cell with a hidden and visible state from two directions (Fig.~\ref{fig:rnn}b), and traverse the lattice in a zig-zag path (Fig.~\ref{fig:rnn}c).
The probability amplitude of the entire configuration $\bm{x}$ with an autoregressive neural-network state is defined as \begin{equation}\label{eq:probrnn}
\psi(\bm{x}) = \prod_{i=1}^N \psi(x_i|x_{i-1},\ldots,x_1).
\end{equation}
The expressivity of this neural network ansatz is determined by the choice of the RNN cell and by the dimension of its hidden state vector $d_h$, also known as the number of hidden units.
The weights of the neural network are updated according to the variational principle: to determine the SCGF in this work, the weights are chosen such that $\langle\psi|W^s|\psi\rangle$ is maximized. 
More details on this optimization are provided in the Supplemental Material (SM).
As the RNN cell itself is not explicitly dependent on the number of lattice sites of the system, it serves as an optimized starting point for further study of large systems: an RNN cell is first extensively optimized on small lattices, which is computationally relatively cheap, after which it can be optimized for a larger system, often requiring only a few hundred iterations until convergence~\cite{Roth2020IterativeNetworks}. 
Hence, the more costly parts of the optimization procedure, such as determining the optimal hyperparameters and avoiding local minima, are only performed for a small lattice, and obtaining results on very large lattices becomes computationally efficient.
Complete architectural and optimization details are provided in the SM.\\

\begin{figure}[t]
\includegraphics[width=\columnwidth]{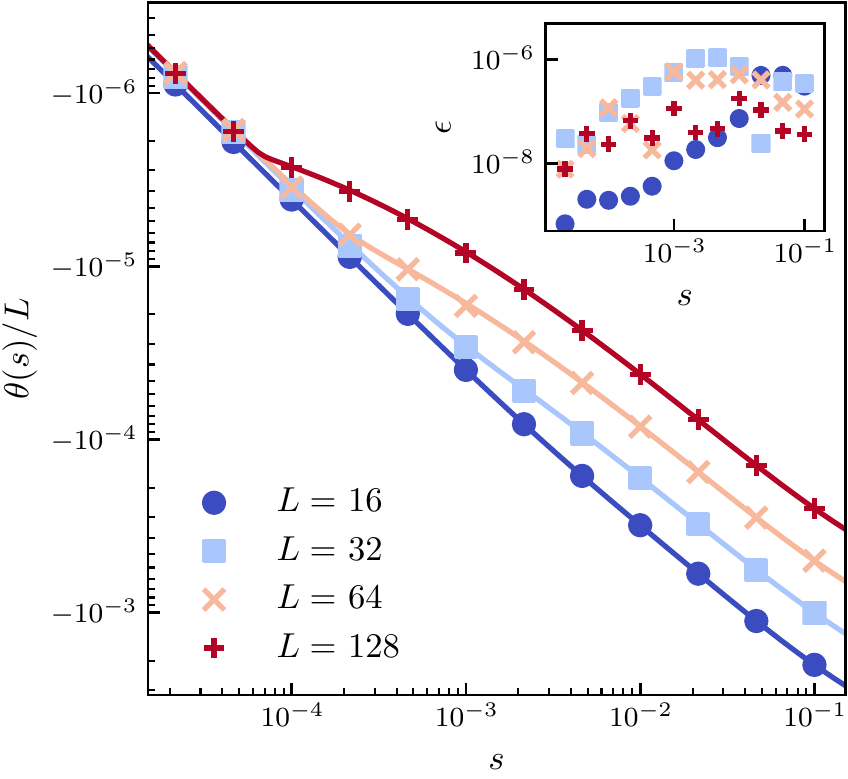} 
\caption{Scaled cumulant-generating function $\theta(s)$ per lattice site of the one-dimensional Fredrickson-Andersen model at $c=0.1$ with a recurrent neural network ansatz (marks) and DMRG (lines) for various values of $s$, and system sizes between $L=16$ and $L=128$. Inset: Difference between DMRG calculations and the RNN ansatz, $\epsilon \equiv \theta(s)_{\text{DMRG}}/L-\theta(s)_{\text{RNN}}/L$. DMRG calculations were performed using the ITensor library~\cite{Fishman2020TheCalculations}.}
\label{fig:1d}
\end{figure}
\textit{1D results}--- We first demonstrate the success of the neural-network state ansatz in computing large-deviation functions by obtaining the SCGF for the dynamical activity of the FA model on a one-dimensional lattice and comparing it to recent results obtained with the density matrix renormalization group (DMRG)~\cite{Banuls2019UsingModels}.
The dynamics described by the tilted generator $W^s$ of the FA model obey detailed balance so that a similarity transformation $P^{-1}W^sP = H^s$ can be performed, with $H^s$ a Hermitian matrix which has the same eigenvalue spectrum as $W^s$. 
The resulting Hermitian matrix is given by~\cite{Garrahan2009First-orderHistories}
\begin{equation} \label{eq:H}
\begin{split}
    H^s = \sum_i &f_i [e^{-s}\sqrt{c(1-c)}\sigma_i^x   \\
    & -c(1-n_i) - (1-c)n_i],
\end{split}
\end{equation}
where $\sigma_x$ is a Pauli matrix.
As $H^s$ is Hermitian, the SCGF obtained through the variational method results in a lower bound on the exact SCGF allowing for a straightforward comparison of the accuracy of the neural-network state ansatz with DMRG.
Here, we use recurrent neural-network states to describe the large deviations of the one-dimensional FA model at $c= 0.1$.
To do so, we optimize RNN states on an $L=16$ chain to find the largest eigenvalue of $H^s$ for values of $s$ corresponding to both the active and inactive dynamical phases. 
Results for larger systems can then efficiently be obtained by using an RNN state optimized at a smaller system size but corresponding to the same dynamical phase; we do so in order to obtain the SCGF for chains with lengths up to $L=128$.
In Fig.~\ref{fig:1d} we compare the SCGF obtained with RNN states to those calculated with DMRG, which reproduce the sharp features characteristic of the model's dynamical phase transition~\cite{Garrahan2007DynamicalGlasses,Garrahan2009First-orderHistories, Garrahan2018AspectsDynamics}. 
The error per lattice site made with the RNN ansatz $\epsilon \equiv \theta(s)_{\text{DMRG}}/L-\theta(s)_{\text{RNN}}/L$ is typically very small, $\epsilon \lesssim \mathcal{O}(10^{-6})$ -- with the largest errors often occurring near the point of largest curvature of the SCGF  -- demonstrating the capability of the RNN ansatz to represent the long-time configurational probability distributions associated with rare trajectories.\\

\begin{figure}[b]
\includegraphics[width=\columnwidth]{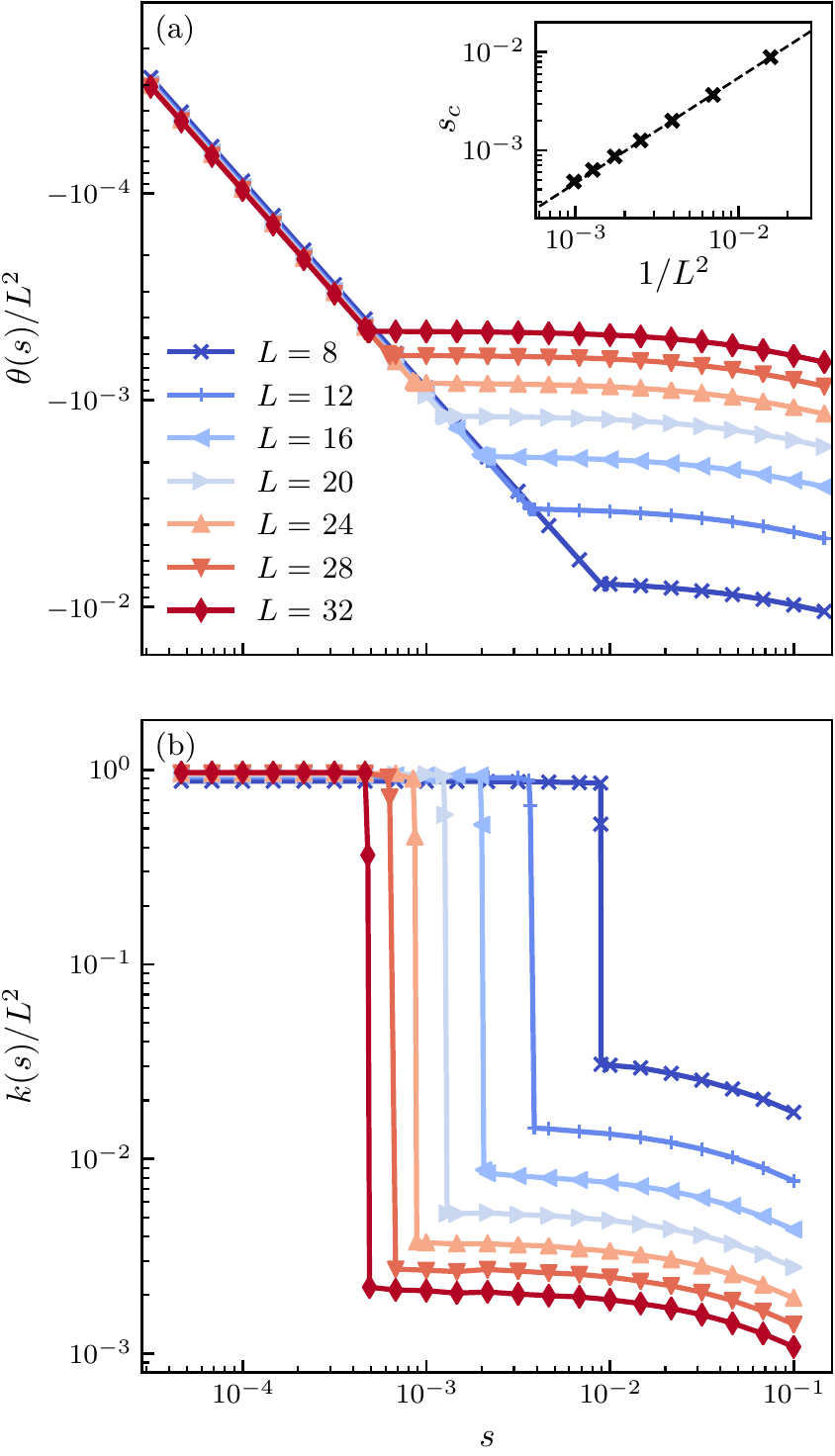} 
\caption{(a) Scaled cumulant-generating function $\theta(s)$ of the two-dimensional Fredrickson-Andersen model at $c=0.5$ with two-dimensional recurrent neural-network states, on $L \times L$ square lattices with length between $L=8$ and $L=32$. Inset: Scaling of the transition point $s_c$ with the number of lattice sites, at which the curvature of $\theta(s)$ is largest. (b) The dynamical activity $k(s) = -\theta^\prime(s)$ per lattice site.
}
\label{fig:2d}
\end{figure}

\textit{2D results}--- 
Having verified the efficacy of the RNN states in computing large-deviation functions, we now turn to the previously unstudied finite-size effects of the FA model in two dimensions. 
To this end we use the two-dimensional RNN shown in Fig.~\ref{fig:rnn}.
Obtaining large-deviation functions in two dimensions with tensor networks has so far been limited to exclusion processes, using either  DMRG~\cite{Helms2019DynamicalStates} or projected entangled pair states (PEPS)~\cite{Helms2020DynamicalNetworks}.
Though shown to be very accurate for two-dimensional quantum systems, the computation of tensor network states for two-dimensional systems is typically expensive, requiring either a large number of variational parameters or scaling unfavorably  with the number of parameters.
Autoregressive neural-network states were recently used to study two-dimensional quantum systems, and have been shown to outperform DMRG~\cite{Hibat-Allah2020RecurrentFunctions} and PEPS~\cite{Sharir2019DeepSystems} for several prototypical models while using far fewer parameters.\\

We first optimize neural-network states for an $8 \times 8$ lattice, at $c=0.5$ and a range of $s$-values, which yields the SCGF $\theta(s)$ (Fig.~\ref{fig:2d}a).
The dynamical activity per lattice site can be calculated as a numerical derivative of the SCGF; $k(s)/N = -\theta^\prime(s)/N$, where $N$ is the number of lattice sites (Fig.~\ref{fig:2d}b). 
Studying the dynamical activity as a function of $s$ reveals a dynamical phase transition at $s_c$, marked by a singularity in the SCGF, between an active phase where $k(s) \approx \mathcal{O}(N)$, and an inactive phase where $k(s)/N\approx 0$, similar to observations in one dimension.
To further characterize this phase transition we calculate how $s_c$ varies with $N$.

Using the RNN states obtained for the $8\times8$ system as starting point, we further optimize neural-network states for a larger system; repeatedly increasing the linear system size by four sites as to obtain the SCGF for system sizes up to $N=1024$.
While the training of the initial RNN state for the $8 \times 8$ system requires $\mathcal{O}(10^4)$ optimization iterations, each successive optimization upon increasing system size converges after $\mathcal{O}(10^2)$ iterations.
The result of this procedure, shown in Fig.~\ref{fig:2d}, reveals that the value of $s_c(N)$ obtained from the peak of the susceptibility $\chi(s)=\theta^{\prime\prime}(s)$ moves toward zero; a finite-size scaling analysis (inset of Fig.~\ref{fig:2d}a) shows that $s_c \sim N^{-\alpha}$ with $\alpha \approx 1.07$. A similar value for this exponent was recently found for the one-dimensional FA model~\cite{Banuls2019UsingModels}.\\

\textit{Outlook}--- We have presented a finite-size scaling analysis of the dynamical activity of the two-dimensional Fredrickson-Andersen model. 
To this end, we have demonstrated the use of the artificial neural-network state ansatz  for obtaining large-deviation functions for classical dynamical systems, drawing from its success in the variational optimization of quantum systems.
Obtaining the scaled cumulant-generating function as an eigenvalue of the tilted generator using this ansatz reveals a transition between phases with high and low dynamical activity at a value of the tilting parameter $s_c$; we have studied the size-dependence of $s_c$.
Our results highlight how the neural-network state ansatz can be employed to efficiently and accurately study large-deviation functions.
Although we have focused our study on a prototypical model, this ansatz is broadly applicable. Given the rapid improvements being made to the neural-network state ansatz, we expect it to play an important role in the study of dynamical large deviations for higher-dimensional systems. \\

\begin{acknowledgments}
Computational resources (Stevin Supercomputer Infrastructure) and services used in this work were provided by the VSC (Flemish Supercomputer Center), and the Flemish Government – department EWI.
T. Vieijra is supported as an `FWO-aspirant' under contract number FWO18/ASP/279. S.W. was supported by the Office of Science, Office of Basic Energy Sciences, of the U.S. Department of Energy under
Contract No. DE-AC02–05CH11231. I.T. acknowledges NSERC. 
\end{acknowledgments}
\bibliography{references}
\clearpage
\newpage

\onecolumngrid
\appendix*

\section*{Supplemental Material}

\subsection*{Variational optimization of the scaled cumulant-generating function}
The expectation value of the tilted generator (Eq.~\ref{eq:H})) can be written as \begin{align*}
\left<\psi|H^s|\psi\right>&= \sum_{x} \left|\psi(x)\right|^2\sum_{x^\prime}H^s_{xx^\prime}\frac{\psi(x^\prime)}{\psi(x)}\\
&= \sum_x \left|\psi(x)\right|^2\theta_{\text{loc}}(x).\\
\end{align*}

The goal of our optimization routine is the maximization of this expectation value, as to obtain the scaled cumulant-generating function $\theta(s)$. 
This is done using variational Monte Carlo: given a parametrized state, $\psi_\mathcal{W}$, we sample the current value of $\left<\psi_\mathcal{W}|H^s|\psi_\mathcal{W}\right> \equiv \Tilde{\theta}(s)$ using $N$ samples $\{x_S\}$ drawn according to $\left|\psi_\mathcal{W}\right|^2$
\begin{equation*}
    \Tilde{\theta}(s) = \frac{1}{N}\sum_{x\in \{x_S\}} \theta_{\text{loc}}(x).
\end{equation*}
The gradients of $\Tilde{\theta}(s)$ w.r.t. to the variational parameters $\mathcal{W}$ are then calculated as
\begin{equation*}
    \partial_\mathcal{W}\Tilde{\theta}(s) = \frac{2}{N}\sum_{x\in\{x_S\}} \partial_\mathcal{W}\log\psi_\mathcal{W}(x) \left(\theta_{\text{loc}}(x) - \Tilde{\theta}(s)\right),
\end{equation*}
and the parameters $\mathcal{W}$ are updated as to maximize $\Tilde{\theta}(s)$.
This is repeated until convergence is achieved, at which point we obtain our best estimate for the SCGF $\theta(s)$.
Determining the SCGF hence requires an efficient way of obtaining uncorrelated samples and evaluating their probability amplitudes.
As explained in the main text, the autoregressive calculation of the probability amplitudes with a recurrent neural network allows for directly obtaining uncorrelated samples in parallel, without the need of Markov chains.\\

\subsection*{Neural network architecture: probability amplitude and sampling}

The recurrent cell used throughout this work is a gated recurrent unit (GRU)~\cite{Cho2014OnApproaches}.
To calculate the conditional probability amplitude $\psi(x_i|x_{i-1},\ldots,x_1)$ at site $i$ of a one-dimensional configuration  $\bm{x} \equiv (x_1,\ldots,x_N)$ of an $N$-site chain, the previous hidden state $h_{i-1}$ (of dimension $d_h$) and spin state $x_{i-1}$ (with local Hilbert dimension $d_s$) are processed through a series of gates to obtain a new hidden state $h_i$:
\begin{align*}
    z_i &= \sigma(W_z x_{i-1} +U_z h_{i-1} + b_u)\\
    r_i &= \sigma(W_r x_{i-1} + U_r h_{i-1} + b_r)\\
    \hat{h}_i &= \tanh(W_h x_{i-1} + U_h(r_i\odot h_{i-1}) + b_h)\\
    h_i &= (1-z_i)\odot h_{i-1} + z_n \odot \hat{h}_i.
\end{align*}
Here, $W$ are $(d_h\times d_s)$ weight matrices, $U$  are $(d_h\times d_h)$ weight matrices, $b$ are biases, $\sigma$ is a sigmoid activation function, and $\odot$ represents the Hadamard product. The update gate $z_i$ calculates how the previous hidden state $h_{i-1}$ is modified by interpolating between $h_{i-1}$ and the candidate hidden state $\hat{h}_i$, and the reset gate $r_i$ determines to what extent the previous hidden state can be ignored. \\

To obtain the conditional probability amplitude $\psi(x_i|x_{i-1}, \ldots, x_1)$, the new hidden state $h_i$ is first passed through a fully-connected layer with $d_s$ output nodes: $y= Wh_i + b$, where $W$ is a weight matrix of dimension $(d_s\times d_h)$ and $b$ is a bias.
A subsequent softmax operation then provides the normalized conditional probability $\Pi_\alpha=e^{y_\alpha}/\sum_{\beta=1}^{d_s}e^{y_\beta}$, where $\alpha$ denotes the index of the local Hilbert space. The conditional probability amplitude is then found as $\psi(x_i|x_{i-1}, \ldots, x_1) = \sqrt{\Pi\cdot x_i}$.\\

To find the total probability amplitude of a given configuration, we start from an initial hidden state $h_0$ and spin $x_0$, which are both set to a zero vector.  
The recurrent cell as described above is then used to traverse the chain and calculates the conditional probability amplitudes at each site.
At the final lattice site visited by the recurrent cell, we enforce the exclusion of the configuration with all spins down from our dynamics. 
To do so, if all prior sites are in the down-state, we enforce the probability of the final site to be 1 for the up-state.
Finally, by multiplying the conditional probability amplitudes for each lattice site, we obtain the total probability amplitude from Eq.~(\ref{eq:probrnn}). \\

Sampling the probability distribution is done in a similar fashion. Again starting from an initial hidden state $h_0$ and spin state $x_0$, we calculate the conditional probability $\Pi$ for site 1. We then sample from this distribution in order to obtain $x_1$, which is then used as input for the next site; this procedure is repeated until $N$ sites have been sampled.\\

For two-dimensional systems, we follow a similar procedure.
In order to respect the geometry of the system, we now propagate hidden states and spin states in both the vertical and horizontal direction (Fig.~1b,c). 
We first process the hidden states $h_\alpha$ and $h_\beta$, coming from the previous sites in the vertical and horizontal direction, to an intermediate hidden state $\Tilde{h} = \Tilde{W}[h_\alpha; h_\beta]$, where $\Tilde{W}$ is a $(d_h \times 2d_h)$ weight matrix acting on the concatenation of $h_\alpha$ and $h_\beta$; and similarly for $x_\alpha$ and $x_\beta$. 
These intermediate hidden and visible states are used as input for a GRU, resulting in a new hidden state and conditional probability amplitude. 
We then traverse the lattice in a zigzag pattern (Fig. 1c), where sites at the edges of the lattice receive zero vectors as previous hidden and spin states.\\

The variatonal ansatz used here allows for a straightforward enforcing of discrete symmetries; here we impose lattice symmetries in the later stages of optimization.
Enforcing such symmetries is achieved as follows: for each configuration $\bm{x}$, we set the probability amplitude as \begin{equation*}
\Tilde{\psi}(\bm{x}) = \frac{1}{N_G}\sum_{\hat{g}\in G} \psi(\hat{g}\bm{x}),
\end{equation*}
where $G$ is the symmetry group with $N_G$ elements.
We use this to enforce parity symmetry for our 1D calculations, and invariance under rotations and reflections for our 2D calculations.
\subsection*{Neural network optimization}

The obtain the SCGFs presented in the main text, we first optimize neural-network states for small lattices ($L=16$ for 1D, $L=8$ for 2D) and a range of $s$-values. In our experiments, we keep the number of hidden units fixed at $d_h = 128$. For each optimization step, we sample 5000 configurations from the current state, and update the neural network weights using Adam~\cite{Kingma2015Adam:Optimization}, with a learning rate of $10^{-4}$, which we decay to $10^{-5}$ until convergence is acquired. We then use these optimized cells as the starting point for a larger system. Should the dynamical phase at a certain $s$ change upon increasing the lattice size, we start from a cell obtained at a nearby value of $s$ which corresponds to the correct dynamical phase. Numerically obtaining the SCGF for very large systems becomes efficient in this way, as each increase in size requires typically only a few hundred training iterations with learning rate $10^{-5}$ until convergence.

\end{document}